\documentclass{camera}
\usepackage{graphicx}  % uncomment this if your want to insert figures

\begin{document}

%%%%%%%%%%%%%%%%%%%%%%%%%%%%%%%%%%%%%%%%%%%%%%%%%%%%%%%%
% The title, only the first letter capitalized; if you want to split it in
% two or more lines, put a \\ macro at each line break
% example: 
%   \title{Title: first line\\ second line}
%
\title{High-energy $\gamma$-ray emission in AGNs and microquasars}
%\title{Microquasars and AGNs as sources of high-energy $\gamma$-rays}
%%%%%%%%%%%%%%%%%%%%%%%%%%%%%%%%%%%%%%%%%%%%%%%%%%%%%%%%
% The author(s), separated by commas; do not put a
% comma before the last author, use instead the \and
% macro which produces a normal ``and'' in the
% caps/small caps context
%
\author{Josep M. Paredes}

%%%%%%%%%%%%%%%%%%%%%%%%%%%%%%%%%%%%%%%%%%%%%%%%%%%%%%%%
%
\organization{Departament d'Astronomia i  Meteorologia, Universitat de Barcelona, Mart\'{\i} i Franqu\`es 1, 08028 Barcelona, Spain}

\maketitle

\begin{abstract} 
	I review here some of the main observational features in microquasars and AGNs as sources of
high-energy $\gamma$-rays as well as some of the current models that try to explain the emission in the
$\gamma$-ray domain.
\end{abstract}

%%%%%%%%%%%%%%%%%%%%%%%%%%%%%%%%%%%%%%%%%%%%%%%%%%%%%%%%
% Write the text starting from here and using the usual
% LaTeX commands.
%
\section{Introduction}

Microquasars are X-ray binary stars  which exhibit relativistic jets, i.e., bipolar outflows of relativistic matter
ejected perpendicularly to both sides of an accretion disc. The jets contain relativistic electrons that
produce the synchrotron radiation detected at radio wavelength. These systems mimic many phenomena seen in
quasars at smaller scales (Mirabel \& Rodr\'{\i}guez 1999). 

The recent detection of TeV $\gamma$-ray emission coming from the microquasars LS 5039 and LS~I~+61~303 (Aharonian et al. 2005a; Albert et al. 2006a), using the ground-based Cherenkov telescopes HESS and MAGIC, is a remarkable result that
consolidate microquasars as a new group of very high-energy (VHE) $\gamma$-ray emitters. Moreover, this
fact strengthes the analogy quasar/microquasar as it is well known that quasars are also sources of
strong $\gamma$-ray emission (Hartman et al. 1999).

\section{AGNs as $\gamma$-ray emitters}

The Third EGRET catalog (Hartman et al. 1999) has shown that active galactic nuclei (AGNs) are strong
$\gamma$-ray (E$>$100 MeV) emitters. At least 30\% of the 271 sources detected above 100 MeV are AGNs and
possibly some more AGNs are present amongst the still unidentified sources (about 60\%). All
high-confidence identified AGNs are blazars and no radio-quiet AGN was detected by EGRET. This
fact can be explained because in blazars there is a highly relativistic outflow of particles that can produce
non-thermal emission extending from radio to $\gamma$-rays, being the high-energy emission produced via
inverse Compton scattering (IC). The relativistic outflow is almost aligned with the observer's line of
sight, producing relativistic beaming and amplifying the flux by several orders of magnitude.

The $\gamma$-ray sky at VHE (E$>$ 100 GeV) contains a few more than 30 sources at the tome of writing (spring of 2006). Most of them are galactic
sources, although there are 12 extragalactic objects (11 blazars and a radio galaxy, M87). Some
of these reported blazars have been seen by HESS (Aharonian et al. 2005b) and/or MAGIC (see Albert et al. 2006b and references therein).
Although by extrapolation from EGRET sources we should expect to detect many more sources, it is necessary
to take into account that VHE $\gamma$-rays
are partially absorbed by the low-energy photons of the extragalactic background light (Aharonian et al. 2006), 
and spectral steepening above GeV energies can occur. 

The spectral energy distribution (SED) of blazars exhibits a double-peaked structure. One peak is in the
optical/X-ray band and is due to synchrotron emission in the magnetic field of the jet. The other peak, located in
the GeV-TeV band, is caused by IC scattering of low-energy photons. These seed photons
can come from outside the jet and produce external Compton scattering (Dermer \& Schlickeiser 1993) or be generated
within the jet as a result of synchrotron radiation (synchrotron self-Compton or SSC) (Maraschi et al. 1992).

\section{How many microquasars do we know?}

Among the known 280 X-ray binaries (Liu et al. 2000, 2001), 43 of them (15\%)  display radio emission 
thought to be of synchrotron origin, being 8 high-mass X-ray binaries (HMXB) and 35 low-mass X-ray
binaries (LMXB) (Rib\'o 2005). At least 15 of them have a relativistic jet and are considered microquasars. This relatively low
number of microquasars probably will increase in the future due to a better sensitivity of the high
resolution radio interferometers, since it might be that the majority of radio emitting X-ray binaries are
microquasars (Fender 2001).

The HMXB microquasars are: LS~I~+61~303, V4641~Sgr, LS~5039, SS~433, Cygnus~X-1 and Cygnus~X-3, and the
LMXB microquasars are: Circinus~X-1, XTE~J1550$-$564, Scorpius~X-1, GRO~J1655$-$40, GX~339$-$4,
1E~1740.7$-$2942, XTE~J1748$-$288, GRS~1758$-$258 and GRS~1915+105. 

Although all of them share the relativistic jet as a common characterisitic, there are some physical
properties that make them individually interesting. Eight of them have (or are suspected to
have) a black hole as a compact object, six of them display superluminal velocities, others have periodic
radio emission, etc. A compilation of the properties of the microquasars in our Galaxy can be found in
Paredes (2005). 

\section{Microquasars as high-energy $\gamma$-ray sources: Theoretical point of view}

The detection of extended non-thermal radio emission from microquasars provided clear evidence for the
presence of relativistic leptons in the jets, although it was not considered that jets could emit
significantly at X-rays or beyond until the detection of X-ray extended emission in some microquasars,
like for instance SS~433 or XTE~J1550$-$564 (Migliari et~al. 2002; Corbel et al. 2002).

Two types of approach have been used to model the emission from jets of microquasars. The hadronic approach
considers that hadrons lead radiative processes at GeV-TeV $\gamma$-rays and beyond, producing significant
amounts of neutrinos, and leaving electrons as possible significant emitters only at lower energies. The
other approach, the so-called leptonic models, extends the energy of leptons from synchrotron radio
emitting energies up to soft $\gamma$-rays exploring several scenarios that included IC and/or synchrotron emission in the jets. 
 
\subsection{Leptonic models}

\begin{figure}
\begin{center}
%\vspace{1cm}
\includegraphics[width=3.9truein]{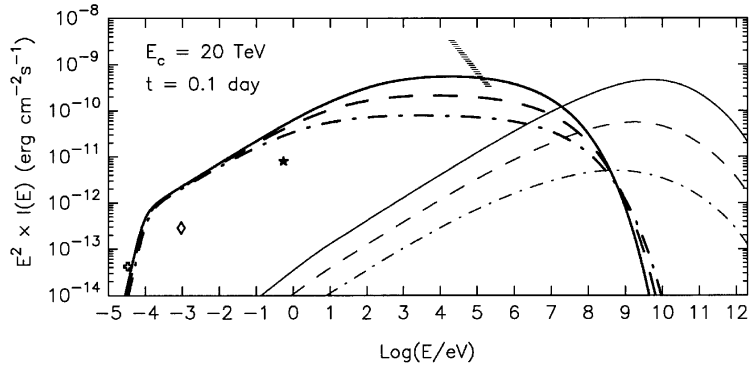}
%\resizebox{\hsize}{!}{\includegraphics{paredes_2006_figrom.ps}}
\caption{The synchrotron (thick lines) and IC (thin lines) emission produced in GRS~1915+105 according to the model of Atoyan \& Aharonian (1999).}
\label{figgr} % optional figure label, must be unique
\end{center}
\end{figure}

One of the firsts leptonic models that were able to produce VHE $\gamma$-rays was developed to explain the
non-thermal flares in GRS~1915+105 (Atoyan \& Aharonian 1999). Relativistic electrons, which were suffering
radiative, adiabatic and energy-dependent escape losses in fast-expanding plasmoids, caused the flares via
synchrotron radiation. If the electrons were accelerated in-situ up to TeV energies, the synchrotron
radiation could then extend beyond the X-ray region,  and the SSC radiation to high and very high energies (see Fig.~\ref{figgr}). In this case, since GRS~1915+105 is a LMXB, the contribution of
the companion star to the (external) IC is not important, while in the case of HMXBs the contribution of
the bright companion can be relevant. Some examples of a leptonic model applied to a HMXB are given by the work
of Kaufman Bernad\'o et~al. (2002) and Georganopoulos et~al. (2002), in which
the authors studied the case where the microquasar jet is exposed to the star photon field. Kaufman
Bernad\'o et~al. (2002)  proposed as well that recurrent and relatively rapid variability could be
explained by the precession of the jet, which results in a variable Doppler amplification. Bosch-Ramon \&
Paredes (2004) developed a leptonic model in which SSC was taken into account,
showing that this mechanism could compete with IC of star photons under certain conditions, and more
recently other authors have developed models of high energy emission from microquasars showing that TeV
photons could be produced (e.g. Bosch-Ramon et~al. 2006, Dermer \& B\"ottcher 2006).

\subsection{Hadronic models}
 
\begin{figure}
\begin{center}
%\vspace{1cm}
\includegraphics[width=2.2truein]{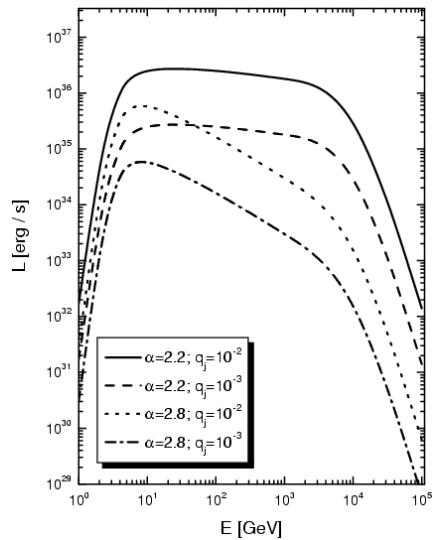}
%\resizebox{\hsize}{!}{\includegraphics{paredes_2006_figrom.ps}}
\caption{Spectral high-energy distribution for models with proton index 2.2 and 2.8, and for different values of the jet/disk coupling parameter $q_{j}$ (Romero et al. 2003).}
\label{figrom} % optional figure label, must be unique
\end{center}
\end{figure}

Several hadronic models to explain the $\gamma$-ray emission from microquasars have been developed in the last
years. One of them, developed by Romero et al. (2003), estimates the $\gamma$-ray emission expected from
the jet-wind hadronic interaction in a high mass microquasar (see Fig.~\ref{figrom}). The $\gamma$-ray emission arises from the decay of neutral pions created in the
inelastic collisions between relativistic protons, moving in a conical jet at energies of $10^{14}$eV, and ions from the strong wind
of the stellar companion. The only requisites for the model are a windy high-mass stellar companion, the presence of
multi-TeV protons in the jet and enough mixing of relativistic protons and wind ions. 
 
%Interactions of hadronic beams with moving clouds in the context of 
%accreting pulsars have been previously discussed in the literature by 
%Aharonian \& Atoyan (1996, Space Sci. Rev. 75, 357). 

Other hadronic models consider a scenario based on high energy protons that escape from the jet and
diffuse through the interstellar medium (ISM) interacting with molecular clouds. Bosch-Ramon et al. (2005)
developed a hadronic model that covers the emission from radio to VHE. They computed the spectrum of 
gamma-rays coming out from the p-p primary interactions via neutral pion decay, as well as the emission
(synchrotron, Bremsstrahlung and IC scattering) produced by secondary particles produced from the 
decay of the charged pions generated in the same p-p collisions. All the relevant energy losses were
taken into account and impulsive and permanent microquasar ejections were considered. 

\section{Microquasars as high-energy $\gamma$-ray sources: Observational point of view}

Microquasars have been targets for most high-energy instruments. In Table~\ref{detections} we present
some of the results obtained using these instruments. In the second column we report count rates obtained with
the IBIS $\gamma$-ray imager on board INTEGRAL, covering the first year of data (Bird et al. 2004). The recent
detection of LS~I~+61~303 in this range of energy has been presented by Chernyakova et al. (2006).

In the third column we have listed the data from BATSE, on board the Compton Gamma Ray Observatory (CGRO),
using the Earth occultation technique (Harmon et al. 2004). Cygnus~X-1 and Cygnus~X-3 have been studied
extensively by BATSE. The microquasars observed by the instrument COMPTEL (Sch\"onfelder et al. 2000), also
on board the CGRO, are listed in the fourth column. The source GRO~J1823$-$12 is among the strongest
COMPTEL sources, and the source region contains several possible counterparts, being LS~5039 one of them
(Collmar 2003). In the case of GRO~J0241+6119, the most likely counterpart is the microquasar LS~I~+61~303,
although its emission in the range 1$-$30~MeV is possibly contaminated by the quasar QSO~0241+622.

\begin{table}[]
\begin{tiny}
%\begin{table}[h!]
\begin{center}
%\begin{table}[!t]
%\begin{small}
\begin{tabular}{lccccc}
\hline
\noalign{\smallskip}Source Name & IBIS/ISGRI & BATSE & COMPTEL & EGRET & CHERENKOV \\
 & 40$-$100 keV & 160$-$430 keV & 1$-$30 MeV & $>$ 100 MeV & TeV \\
 & (count/s)    & (mCrab)       &  (GRO)     & (3EG)      &      \\
\hline
\noalign{\smallskip}
\multicolumn{6}{c}{\bf High Mass X-ray Binaries (HMXB)}    \\
\hline
\noalign{\smallskip}
% \tablehead{1}{l}{b}{Name\\  \\} &
% \tablehead{1}{c}{b}{INTEGRAL\tablenote{The first IBIS/ISGRI 
% soft $\gamma$-ray galactic plane survey catalog 
%(\cite{bird1}).}\\40$-$100 keV\\(count/s)} &
% \tablehead{1}{c}{b}{BATSE\tablenote{BATSE Earth occultation catalog, Deep 
% sample results 
%(\cite{harmon1}).}\\160$-$430 keV \\   (mCrab)}  &
% \tablehead{1}{c}{b}{COMPTEL\tablenote{The first COMPTEL source catalogue 
% (\cite{schonfelder1})}\\
%  1$-$30 MeV \\ (GRO) } &
% \tablehead{1}{c}{b}{EGRET\tablenote{The third EGRET catalog of 
% high-energy $\gamma$-ray sources (\cite{hartman1})}\\ $>$ 100 MeV \\ (3EG)} %\\
%\hline
%\tablehead{5}{c}{b}{\bf High Mass X-ray Binaries (HMXB)} \\
%\hline

LS~I~+61~303 &  yes  & 5.1$\pm$2.1  & J0241+6119? & J0240+6103 & MAGIC~J0240+6115\\

V4641~Sgr  & $-$ & $-$   & $-$ & $-$ & $-$\\

LS~5039 &  $-$ &  3.7$\pm$1.8 & J1823$-$12?  & J1824$-$1514 & HESS~J1826$-$148\\
  
SS~433 & $<$1.02  &  0.0$\pm$2.8 & $-$ & $-$  &$-$\\
  
Cygnus~X-1  & 66.4$\pm$0.1  & 924.5$\pm$2.5 &  yes &$-$ & $-$\\
  
Cygnus~X-3    &  5.7$\pm$0.1     & 15.5$\pm$2.1 & $-$   & $-$ &$-$\\

\hline
\noalign{\smallskip}
\multicolumn{6}{c}{\bf Low Mass X-ray Binaries (LMXB)}    \\
%\tablehead{5}{c}{b}{\bf Low Mass X-ray Binaries (LMXB)}\\
\hline
\noalign{\smallskip}
      
Circinus~X-1    &  $-$     &   0.3$\pm$2.6&  $-$   & $-$ &$-$\\
 
XTE~J1550$-$564 &  0.6$\pm$0.07    & $-$2.3$\pm$2.5 & $-$   & $-$ &$-$\\
 
Scorpius~X-1     &  2.3$\pm$0.1      &  9.9$\pm$2.2 &  $-$ & $-$ & $-$\\
  
GRO~J1655$-$40 & $-$    &   23.4$\pm$3.9 & $-$ & $-$ & $-$\\
   
GX~339$-$4   &   0.55$\pm$0.03    &   580$\pm$3.5 & $-$&$-$ & $-$\\ 
  
1E~1740.7$-$2942&  4.32$\pm$0.03 &  61.2$\pm$3.7 & $-$& $-$ &$-$ \\
 
XTE~J1748$-$288 &   $-$  &    $-$ & $-$ & $-$      & $-$ \\

GRS~1758$-$258  &  3.92$\pm$0.03  &  38.0$\pm$3.0  & $-$ &$-$\ &$-$\\
   
GRS~1915+105    &   8.63$\pm$0.13 &  33.5$\pm$2.7 & $-$  & $-$ &$-$\\
\hline
\end{tabular}
\caption{High-energy emission from microquasars.}
\label{detections}
%\end{small}
\end{center}
\end{tiny}
\end{table}

\begin{figure}
\begin{center}
\includegraphics[width=3truein]{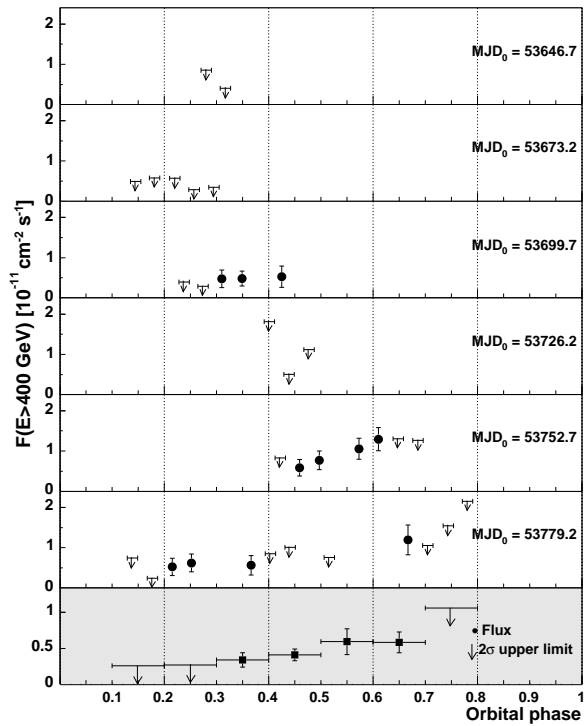}
%\resizebox{\hsize}{!}{\includegraphics{paredes_2006_figlsi.eps}}
\caption{MAGIC observations of LS~I~+61~303 (Albert et al. 2006a), showing hints of a periodic nature of the VHE emission. Periastron is at phase 0.23}
\label{figlsi} % optional figure label, must be unique
\end{center}
\end{figure}

In the fifth column there are the two microquasars associated to EGRET sources. The discovery of the
microquasar LS~5039 and its association with 3EG~J1824$-$1514 (Paredes et al. 2000) provided observational
evidence that microquasars could also be sources of high-energy $\gamma$-rays. The radio source
LS~I~+61~303 was historically associated with the $\gamma$-ray source 2CG~135+01 (Gregory \& Taylor, 1978)
and later with the EGRET source 3EG~J0241+6103. After the discovery of relativistic jets in LS~I~+61~303
(Massi et al. 2001, 2004) the source was considered the second microquasar associated with an EGRET source.
The association of these two microquasars with high-energy $\gamma$-ray sources has been confirmed after
the detection of LS~5039 by HESS (Aharonian et al. 2005a) and of LS~I~+61~303 by MAGIC (Albert et al. 2006a),
as can be seen in the sixth column. In Fig.~\ref{figlsi} we show the VHE $\gamma$-ray flux of
LS~I~+61~303 as a function  of the binary system orbital phase (orbital period of 26.496 days) for the six observed
orbital cycles. These are the first observational evidences of variability (and possibly periodicity) of the VHE $\gamma$-ray emission in microquasars.

\section{Two striking microquasars: LS~5039 and LS~I~+61~303}

After the suggestion that LS~5039 and LS~I~+61~303 were the radio counterparts of two EGRET sources,
several specific models aimed to explain the observed HE emission and to predict the level of VHE emission
were developed. After the detection of TeV emission from both sources, more detailed models have been
developed to explain the high-energy and VHE spectrum. It is worth noting that any TeV emission model must
take into account the strong opacity effects on the $\gamma$-rays introduced by different photon fields,
which are particularly strong in massive binaries. 

\begin{figure}
\begin{center}
\includegraphics[width=3.3truein]{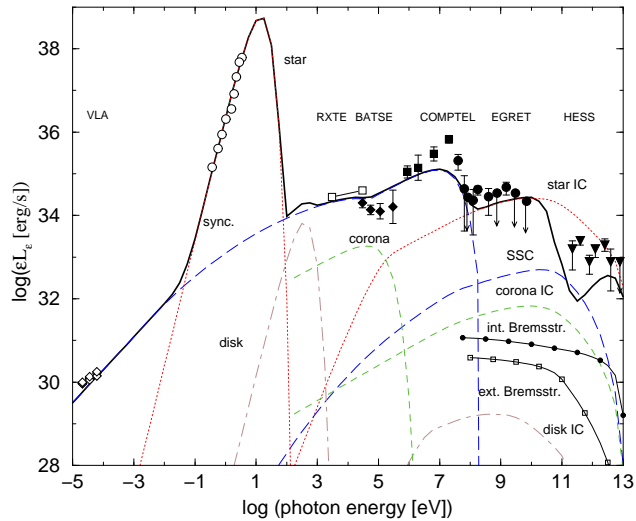}
%\resizebox{\hsize}{!}{\includegraphics{paredes_2006_figls.eps}}
\caption{SED of LS~5039. The main radiation components are shown, altogether with the total emission, which suffers attenuation by photon photon absorption above 10 GeV (Paredes et al. 2006).}
\label{figls} % optional figure label, must be unique
\end{center}
\end{figure}

In the context of microquasars, Bosch-Ramon et al. (2006) developed a model based on a freely expanding
magnetized jet, in which internal energy is dominated by a cold proton plasma extracted from the accretion
disk. The model was applied to LS~5039 (Paredes et al. 2006) to reproduce qualitatively the spectrum of
radiation produced in the jet from radio to VHE. In
Fig.~\ref{figls} we show the SED predicted for LS~5039 and the data obtained with several instruments. The effects of the absorption are evident in the SED, where
there is a minimum around few 100~GeV (see Fig.~\ref{figls}).  

In the case of LS~I~+61~303, Romero et al. (2005) presented a hadronic model for $\gamma$-ray production
where they calculated the $\gamma$-ray emission originated in p-p interactions between relativistic protons
in the jet and cold protons from the wind. The model takes into account the opacity of the ambient photon
fields to the propagation of the $\gamma$-rays. 

The propagation of VHE $\gamma$-rays inside LS 5039 and LS~I~+61~303 has been studied recently by Bednarek
(2006). Since the massive stars in these systems are very luminous, the high-energy $\gamma$-rays injected
relatively close to the massive stars should be strongly absorbed, initiating electromagnetic
cascades. A part of the primary $\gamma$-rays and secondary cascade
$\gamma$-rays escape from the binary system toward the observer. The cascade processes occurring inside
these binary systems significantly reduce the $\gamma$-ray opacity obtained in other works by simple
calculations of the escape of $\gamma$-rays from the radiation fields of the massive stars. The maximum in
TeV $\gamma$-ray light curve predicted by the propagation effects in LS~I~+61~303 should occur after
periastron passage (as has been observed). 

Maraschi \& Treeves (1981) proposed an scenario based on the interaction of the relativistic wind from a young non-accreting pulsar with
the wind from its stellar companion to explain the high-energy observations of LS~I~+61~303. Dubus (2006) has used this scenario to model the overall SED of LS~5039 and LS~I~+61~303. Although it can explain some characteristics of these
sources, there are several properties that are difficult to be explained: the jet shape and the persistency in its direction in LS~5039, the radio outbursts in LS~I~+61~303, the fast precessing radio jets in LS~I~+61~303 and the GeV $\gamma$-ray fluxes in both sources.

\section{Summary}
Up to now there are 66 EGRET blazars and 11 TeV blazars detected. On the other hand, microquasars have consolidated as a new source population of the $\gamma$-ray sky. As emitters of HE
and VHE photons, they also mimic the high-energy behaviour of quasars. This fact puts microquasars among the
most interesting sources in the Galaxy from the viewpoint of high-energy astrophysics.

Several kind of models predict that radio jets could be natural sites for the production of high-energy
photons via both Compton scattering and proton proton collisions in microquasars. 

Additional microquasars will very likely be detected soon with the Cherenkov telescopes and GLAST. This
will bring more constraints to the physics of these systems and their relationship with AGNs.

\section{Acknowledgements}
I acknowledge partial support by DGI of the Ministerio de Educaci\'on y Ciencia (Spain) under grant
AYA2004-07171-C02-01, as well as additional support from the European Regional Development Fund
(ERDF/FEDER).  I thank Marc Rib\'o, Valent\'{\i} Bosch-Ramon and Josep Mart\'{\i} for a careful reading of the manuscript and useful comments.

%%%%%%%%%%%%%%%%%%%%%%%%%%%%%%%%%%%%%%%%%%%%%%%%%%%%%%%%
% End of the paper
%
\end{document}